
\input phyzzx

\def\a{\rightarrow}
\def\aa{\alpha}
\def\b{\beta}
\def\d{\partial}
\def\({\left (}
\def\[{\left [}
\def\){\right )}
\def\]{\right ]}
\def\R{\tilde R}
\def\rt{\tilde{\rho}}
\def\rd{\rho_{dust}}
\def\rc{\rho_{charge}}
\def\vx{\vec{x}}
\def\vr{\vec{r}}
\def\vri{\vec{r}_i}
\def\tout{\theta_{out}}
\def\tin{\theta_{in}}

\REF\pen{R. Penrose, {\it Rel. del Nuovo Cimento} {\bf 1}, 252 (1969).}
\REF\wald{R.M. Wald, {\it Ann. Phys.} {\bf 82}, 548 (1974).}
\REF\kt{D. Kastor and J. Traschen,
{\it Phys. Rev. } {\bf D47}, 4476 (1993).}
\REF\cauchy{S. Chandrasekhar and J. Hartle, {\it Proc. Roy. Soc.
London} {\bf A384}, 301 (1982); E. Poisson and W. Israel, {\it Phys. Rev.}
{\bf D41}, 1796 (1990); A. Ori, {\it Phys. Rev. Lett.} {\bf 67}, 789 (1991).}
\REF\dscauchy{
F. Mellor and I. Moss, {\it Phys. Rev.} {\bf D41},
403 (1990).}
\REF\scri{T. Damour, in Proceedings of the Fourth Marcel Grossman Meeting,
ed. R. Ruffini (North-Holland, Amsterdam, 1986) and references therein.}
\REF\snkm{T. Shiromizu, K. Nakao, H. Kodama and K. Maeda,
{\it Phys. Rev. }{\bf D47}, R3099,
(1993)}
\REF\bh{D. Brill and S. Hayward, {\it Global Structure of a Black Hole Cosmos
and Its Extremes}, preprint gr-qc/9304007 (1993).}
\REF\mp{S.D. Majundar, {\it Phys. Rev.} {\bf 72}, 930 (1947),
A. Papapetrou, {\it Proc. Roy. Irish Acad.} {\bf A51}, 191 (1947).}
\REF\nym{K. Nakao, K. Yamamoto and K. Maeda,
{\it Phys.~Rev.} {\bf D47}, 3203 (1993)}
\REF\piotr{P.T. Cru\'sciel and D.B. Singleton,
{\it Commun.~Math.~Phys.} {\bf 147}, 137 (1992)}
\REF\schmidt{B. Schmidt, private communication}
\REF\hh{J.B. Hartle and S.W. Hawking, {\it Comm. Math. Phys.} {\bf 26}, 87
(1972).}
\REF\bl{D.R. Brill and R.W. Lindquist, {\it Phys. Rev.} {\bf 131},
471 (1963)}
\REF\boulware{D. Boulware, {\it Phys. Rev.} {\bf D8}, 2363 (1973).}

\Pubnum={NSF-ITP-93-78\cr  UMHEP-387 \cr gr-qc/9307014}
\titlepage
\title{Testing Cosmic Censorship with Black Hole Collisions}
\author{Dieter R. Brill and Gary T. Horowitz}
\address{Institute for Theoretical Physics\break
	 University of California \break
	 Santa Barbara, CA. 93106}
\author{David Kastor and Jennie Traschen}
\address{Department of Physics and Astronomy\break
University of Massachusetts\break
	 Amherst, Massachusetts 01003}
\vfil
\abstract

There exists an upper limit on the mass of  black holes when the
cosmological constant $\Lambda$ is positive. We study the collision of
two black holes whose total mass exceeds this limit.
Our investigation is based on a recently discovered exact solution describing
the collision of $Q=M$ black holes with $\Lambda > 0$. The global structure
of this solution is analyzed. We find that if the total mass is less than
the extremal limit, then the black holes coalesce. If it is greater, then
a naked singularity forms
to the future of a Cauchy horizon. However, the horizon is not smooth.
Generically, there is a mild curvature singularity, which still allows
geodesics
to be extended. The implications of these results for cosmic censorship
are discussed.

\endpage
\chapter{Introduction}

It is almost 25 years since Penrose first proposed his cosmic censorship
conjecture [\pen]. Roughly speaking, this conjecture states that naked
singularities do not form from
realistic physical processes. This conjecture is widely believed to be true
and has become the cornerstone
of our understanding of gravitational collapse and black holes. But
despite extensive work over the years, we are still far from having a proof.
One appears to need global existence results for strong-field,
nonsymmetric solutions, which are extremely difficult to obtain.

When a proof seems difficult, it  may be easier to obtain a
convincing counterexample and show that the conjecture is false.
Many attempts have been made. One of the earliest is the following.
 Static charged black holes in the
Einstein-Maxwell theory are characterized by their mass $M$ and charge $Q$,
and exist only
for $Q \le M$. For $Q>M$ the spacetime describes a naked singularity.
One can ask whether it is possible to start with a black hole with $Q=M$,
drop in a test particle with charge greater than its mass, $q > m$,
and turn the black hole into a naked singularity. Wald showed
that this cannot happen [\wald].
In order that a $q>m$ test particle may reach the
horizon, it must be thrown in with sufficient kinetic energy that the mass
of the black hole increases more than its charge.

In the presence of a positive cosmological constant $\Lambda$,
there is a variation
of this test. Charged black holes in a de Sitter background have three
horizons,
an inner and outer black hole horizon, and a de Sitter horizon.
There are thus two extremal limits. One corresponds to coinciding inner and
outer black hole horizons, and is analogous to the extremal
Reissner-Nordstr\"om solution.
The other corresponds to coinciding outer black hole and de Sitter
horizons. In both cases, exceeding these extremal cases results
in naked singularities.
For the first type of extremal
limit, it was shown in [\kt] that it is again impossible to destroy the black
hole by sending in charged test particles.  However,
we will show in Sec. 2 that for the second type
of extremal black hole, one {\it can} drop in a charged
test particle and exceed the limiting value.
One thus seems to have a potential
violation of cosmic censorship when $\Lambda >0$.

To analyze this situation further, one needs to go beyond the test particle
approximation and include the backreaction of the test particle on the
geometry. Ideally one would like to have an exact solution describing
two black holes colliding in de Sitter space. Remarkably,  a solution of
exactly this type has recently been found [\kt]. The solution describes an
arbitrary number of $Q=M$ black holes with $\Lambda >0$. The solution is
dynamical, and the black holes collide in the future. (Alternatively,
one can consider the time reverse, which describes white holes
splitting and expanding.)
For small $M$,
the single $Q=M$ black hole
is not extremal and has three horizons. However, when one
increases $M$, there is an extremal value
where the outer black hole and de Sitter
horizons coincide. One can thus consider two black holes each with mass
less than this extremal value, but whose sum is greater, and let them collide.
The purpose of this paper is to examine this process and
determine whether cosmic censorship is indeed violated.

In order to settle this question, several preliminary steps are necessary.
Some of these have interesting applications independent of cosmic censorship.
First, one needs to investigate the global
structure of the multi-black-hole solutions. This requires extending
beyond the coordinate patch in which the solutions were first presented.
Here we find a surprise.
The horizons across which one must extend are in general not smooth,
but have only finite differentiability. As a result, the extensions are
not unique.
This has
implications for the instability of the inner black hole horizon
[\cauchy, \dscauchy], and yields a
possible analogy to the lack of smoothness of null infinity in asymptotically
flat spacetimes [\scri].
Physically, for certain horizons, the lack of smoothness
can be interpreted as the result of electromagnetic and gravitational
radiation in the spacetime which is not smooth at the horizon.
We also study null geodesics and determine the global event
horizon. We find that when the total mass is less than the extremal limit,
the black holes coalesce.

Since $\Lambda >0$,  these solutions are not asymptotically flat, and
the usual definitions of black holes and naked singularities are no
longer valid. However, for spacetimes that asymptotically approach
de Sitter, there is an analog of future null infinity, and one can define
the event horizon to be the boundary of the past of this null infinity. Since
the singularities of a charged black hole are timelike, they are always
locally naked. However, if they are enclosed inside an event horizon, there
is no violation of weak cosmic censorship. We will be interested in
whether the singularity is visible to {\it all} observers who start from
a given asymptotic de~Sitter region in the past. If so, there would be
no way for these observers to avoid seeing the singularity, and even
weak cosmic censorship would be violated.

Several properties of black holes in de~Sitter space have recently been studied
by Shiromizu et. al. [\snkm].
They discuss black hole collisions, but without the benefit of
an exact solution. By assuming cosmic censorship, they are lead to the
conclusion that large black holes will not collide when $\Lambda > 0$.
We will see that the exact solution behaves quite differently.

By studying this solution, we find that the question of whether
cosmic censorship is violated by charged black hole collisions is rather
subtle. Three different issues must be dealt with.
First, since the  total mass exceeds the extremal limit, there
is always a naked singularity  ``on the other side" of de Sitter space,
even before the black holes collide. Thus, one is essentially starting
with singular initial conditions and should not be surprised if
cosmic censorship is violated. We will show that this problem can be
alleviated by introducing a charged shell of dust, which will remove the
unwanted singularity.
Second, there is always a Cauchy horizon,
which we will show is generically singular.
However, the singularity is rather mild, and  geodesics can
pass through it. Furthermore, there is a large subset
of the initial data
for which the curvature singularity is removed. In these cases, all
observers cross the Cauchy horizon and see a naked singularity. Cosmic
censorship is violated. However, the known exact solutions do not
describe the most general black hole collision with a positive cosmological
constant. It is certainly possible that in the generic collision
the singularity at the Cauchy horizon will be
much worse, and there will be no way to continue the spacetime.
Finally,
the naked singularity appears only for eternal black holes. We will see that
one can form these black holes from regular initial conditions (e.g. using
charged dust). But in this case,  the matter forming the black holes
and the matter introduced to remove the unwanted singularity collide
before a naked
singularity is reached. We do not have the exact solution beyond this
point, and so do not know if naked singularities will form in the future.

Analysis of the global structure of the multi-black-hole solutions is
complicated by the fact that, in general, they are neither static, nor
spherically symmetric. In light of this, we begin in Sec.~2 by discussing
the special case of a single black hole. This is described by the
Reissner-Nordstr\"om-de Sitter solution. We will show that, in the test
particle approximation, a $Q=M$ black hole can have its mass
increased past the extremal
limit of coinciding black hole and de Sitter horizons.
In Sec.~3 we start our discussion of the multi-black-hole solutions
by examining the basic features of their geometry. By locating the trapped
surfaces, we see here the first
evidence that some of them describe coalescing black holes.
In Sec.~4 we investigate extensions of the solutions and
discuss the lack of smoothness at the horizons. Sec.~5 contains a discussion
of the event horizon and establishes that (when the total mass is less
than the extremal limit) its topology indeed changes,
showing that black holes do combine. In Sec.~6
we consider the introduction of charged dust. Finally, in Sec.~7, the
possible counterexample to cosmic censorship is studied. Sec.~8 contains
some concluding remarks.

\chapter{The $Q=M$ Reissner-Nordstr\"om-de Sitter Solution}

The Reissner-Nordstr\"om-de Sitter (RNdS)
solution is static and spherically symmetric,
and its global structure can be analyzed by general methods.  This has been
done by Brill and Hayward [\bh]. Here we will briefly recount
their results for the case $Q=M$.  We also discuss briefly the motion of
test particles in $Q=M$ RNdS backgrounds.  We show that one can add
$q=m$ test particles to an extremal black hole, causing it to exceed the
extremal limit.

\section{Static Coordinates}
The static form of the $Q=M$ RNdS metric and electromagnetic potential is
$$ ds^2 = -V(\R)dT^2 + {d\R^2\over V(\R)} + \R^2d\Omega^2 ,$$
$$V(\R)=\left( 1-{M\over \R}\right ) ^2 - {\Lambda\over 3}\R^2, \qquad
A_T = -{M\over \R}. \eqn\static $$
We assume $\Lambda >0$, and will interpret $M$ as the mass of the black hole.
There is a curvature singularity at $\R=0$.  Horizons occur where $V(\R)=0$ and
$\R>0$.  The number of horizons depends on the mass and cosmological constant.
For $M^2\Lambda < {3\over 16}$, there are three horizons: inner and outer black
hole horizons and a de Sitter horizon.  If the mass of the black hole is
increased, the outer black hole and de Sitter horizons move closer
together. They coincide at the extremal mass given by
$$M_{ext}^2 \equiv {3\over 16 \Lambda}. \eqn\extl $$
For $M>M_{ext}$ there is only a single horizon.
We will call this the ``overmassive" case.

The Penrose diagrams for these three cases are shown in Figs. (1 -- 3).
Fig. (1) shows the case $M<M_{ext}$ with its three horizons. Notice that
$\Im^\pm$ are spacelike when $\Lambda >0$.
The overmassive case is shown in Fig.  (3), and clearly has naked
singularities.
However, these singularities exist for all time, and hence this is not
a violation of cosmic censorship.
The extremal limit, Fig. (2), is rather unusual. This solution
has singularities, which are visible from
$\Im^+$, and also a nonsingular spacelike
surface to their past.
(By making an appropriate identification this surface can even be
compact.) However, there are complete timelike geodesics along which
the singularity is never visible. (These reach the point $p$ in the figure.)
Thus, one can view the point $p$ as a
future timelike infinity and define a horizon to be the boundary of its
past. This illustrates the difficulty of defining a naked singularity
when the cosmological constant is nonzero. As stated earlier, we will call
a singularity naked if it is visible to {\it all} observers originating from a
given asymptotic de Sitter region in the past. So the extremal limit
is not naked in our sense.

\section{Cosmological Coordinates}
The $Q=M$ RNdS solution can also be expressed in isotropic or
``cosmological" coordinates:
$$ds^2= -{d\tau^2\over U^2} +U^2 \left( dr^2 +r^2 d\Omega ^2\right ),\qquad
U=H\tau + {M\over r}, \qquad A_\tau = {1\over U}. \eqn\cosmo $$
where $H =\pm  \sqrt{\Lambda/3}$.
The transformation from the static coordinates to cosmological coordinates
is given by
$$\R = H\tau r +M,\qquad T={1\over H}\log H\tau -h(\R),\qquad
{d h \over d \R} =-{H\R^2\over (\R-M)V(\R)}.\eqn\transform $$
Since \cosmo\
is the form of the solution that is easily generalized to several
masses,
it is useful to understand how these
cosmological coordinates cover the spacetime.
In Fig. (1), a single patch of $(r,\tau)$ coordinates with $H<0$ covers the
region enclosed by the bold lines.  Some constant-$\tau$ surfaces have been
drawn in this region.  Below the dashed line $\tau$ is negative.  The
left-hand end of the constant $\tau$ surfaces is $r=\infty$.  The right-hand
end is $r=0$ (but we will soon see that
this is an ``infinite throat," not a regular origin
of polar coordinates).  Above the dashed line  $\tau$ is positive.
In this region, the right-hand end of the
constant-$\tau$ surfaces is still $r=0$, but the chart ends on the left
at a finite
$r$, where $U$ vanishes and the geometry is
singular.  Note that this coordinate patch covers a past de Sitter
horizon and the inner and outer black hole horizons. The future de Sitter
horizon corresponds to $r = \infty,~ \tau = 0$, and the region beyond
this horizon, where $\Im^+$ lies, is not covered by these coordinates,
but it is covered by \cosmo\ with $H > 0$.

The extremal limit and overmassive cases, Figs. (2) and (3), are similar.
Again we have drawn some constant-$\tau$
surfaces and again $\tau =0$ is shown as a dashed line.  The geometry of the
spatial slices for $\tau <0$ is regular, but for $\tau >0$, $r$  again
extends only to a finite maximum value at the singularity.

\section{Paths of Test Particles}

Consider a $q=m$ test particle with conserved energy $E=m$ moving
in a $Q=M$ RNdS spacetime. It was shown in [\kt] that the radial motion,
in static coordinates, is given by
$${d\R\over d\lambda}=\pm H\R,\qquad {dT\over d\lambda}=\pm
{1\over V(\R)}\left (1-{M\over\R}\right ) , \eqn\testpath  $$
where $\lambda$ is the proper time along the path.  There are two possibilities
for the motion depending on whether the signs in \testpath\ are chosen
to be the
same or opposite.  The other choices of sign in \testpath\ reverse the flow of
time along these paths.  It is simple to check, using \transform ,
that choosing both plus signs in \testpath\ gives a path of constant $r$
in cosmological coordinates.  Similarly, we can define a new set of
cosmological coordinates $(r^\prime,\tau ^\prime)$ by reversing the sign of
the static time coordinate $T$ in \transform .  The paths \testpath ,
with opposite choices of signs are then paths of constant $r^\prime$.
These paths of constant $r^\prime $ will be important to us in Sec. 6 when
we discuss the collapse of charged shells.

Now consider a test particle on a path of constant $r$, such as the one shown
in Fig. (1).
At early times the test particle is
outside both the de Sitter and black hole horizons.  As time progresses, it
enters first the de Sitter and then the black hole horizon.  Nothing
in the above discussion relied on the black hole being undermassive.
If the black
hole is an extremal one, the description is roughly the same, with the test
particle passing through the degenerate horizon.
It follows immediately that
there is no barrier to such a test particle entering the horizon.
One can exceed the extremal mass limit by dropping in test particles.
Furthermore, since the charges on the black hole and test particle provide a
repulsive force, we expect that this phenomenon will be generic for
black
holes and test particles with charge less than their mass.

\chapter{General Properties of the Multi-Black-Hole Solutions}
\section{Geometry of spatial surfaces}
The solutions of reference [\kt] depend
on a number of parameters that correspond to several
different masses at arbitrary positions (but not arbitrary velocities).
The metric and
gauge field for mass parameters $M_i$ and positions
$\vec{r}_i$ are given by
$$\eqalign{ds^2 & =-{d\tau^2\over U^2} + U^2 d\vec{r}\cdot d\vec{r},
\qquad A_\tau ={1\over U} \cr
U & = H\tau+\sum_i {M_i\over |\vec{r}-\vec{r}_i|}, \qquad\qquad
H=\pm\sqrt{\Lambda/3}. }\eqn\metric $$
{\it Unless otherwise stated, we will assume $H<0$.} It is only for
this case that the spacetime describes black hole collisions.
In terms of $H$, the extremal limit \extl\ is
$$  M_{ext} = {1\over 4 |H|}. \eqn\extremal  $$

The surfaces of constant $\tau$ are spacelike everywhere. Near each
$\vr_i$ the geometry resembles the infinite throat familiar from
the asymptotically flat extremal Reissner-Nordstr\"om solution.
This can be seen by
expressing the spatial metric in spherical coordinates centered
at $\vr =\vri $. Near the origin of these coordinates, this metric becomes
$$dl^2 \approx {M_i^2\over r^2}dr^2 +M_i^2 d\Omega ^2, \eqn\throat $$
which is the metric for a cylinder of infinite spatial extent having
cross sectional area $4\pi M_i^2$.

The curvature of \metric\ can be singular
at zeros of the metric function $U$.
This can be seen from the square of the Maxwell field strength,
$$ F^2 \equiv F_{\mu\nu}F^{\mu\nu}=
-{(\vec{\nabla} U)^2\over U^4}\eqn\maxwell $$
where $\vec{\nabla}$ denotes the usual gradient with respect to $\vr$.
If $U=0$ and
$\vec{\nabla} U$ does not vanish like $U^2$ or faster, then $F^2$
diverges and the curvature is singular. The metric is
regular everywhere else, but it is incomplete as discussed in Sec. 4.

Assembling these elements we get a picture of how the spatial
geometry develops in cosmological time. We will start with one black hole
first.
Consider the spatial surfaces for a single
black hole with mass $M<M_{ext}$, which are sketched on Fig. (1).
The metric function is $U = H\tau +{M\over r}$.  For $\tau<0$,
$U$ is positive everywhere and the spatial surfaces are nonsingular. They
are asymptotically flat
and have the cylindrical form of an infinite throat near the origin.
For $\tau=0$, the spatial metric is regular and has the cylindrical form
everywhere.  As $\tau$ is
increased slightly, a singularity appears near $r =\infty$. As $\tau$ increases
further, the singularity cuts off more and more of the cylinder. In Fig. (1)
this is shown by the way the singularity intersects the
spatial surfaces.

The generalization to more than one mass is then straightforward.
For $\tau <0$ the spatial surfaces are nonsingular and asymptotically flat at
large radius.  Near each ``point" $\vri$, the spatial metric has
the form \throat\ of a throat.  These surfaces are depicted in Fig. (4a).
For $\tau =0$, the surface is nonsingular, but spatial infinity is now also
asymptotically cylindrical.  This surface is depicted in Fig. (4b).
As $\tau$ is increased slightly above zero, a singularity moves in from
spatial infinity as shown in Fig. (4c).
As $\tau$ continues to increase, the singularity splits and eventually
surrounds each of the throats individually.
The spatial surface is then composed of a number of isolated throats, as
depicted in Fig. (4d).
This description is independent of the size of the masses, and hence applies
to both the overmassive and undermassive cases.

\section{Trapped Surfaces}
The causal structure of the multi-black-hole solutions is rather involved.
To gain some understanding, we begin by considering
trapped surfaces.
The expansions $\theta_{out}$ ($\theta_{in}$) of outgoing (ingoing) null rays
normal to a 2-surface $\Sigma$ in a spatial hypersurface are given by
$$\theta = D_i n^i -K_{ij}n^i n^j + K, \eqn\expansion $$
where $h_{ij}, K_{ij}$ are the metric and extrinsic curvature on the spatial
hypersurface, $D_i$ is the covariant derivative compatible with $h_{ij}$,
$K = h_{ij}K^{ij}$,
and $n^i$ is the outward (inward) directed normal vector to
$\Sigma$.
A surface $\Sigma$ is called outer trapped if $\theta_{out}<0$, and
inner trapped if $\theta_{in}>0$.
Surfaces for
which $\tout$ ($\tin$) vanish are called outer (inner) apparent horizons.
For the surfaces of constant
$\tau$ in the metric \metric, the extrinsic curvature is simply
given by $K_{ij}=H h_{ij}$.

In the case of a single black hole of mass $M$, one can calculate the
expansions
exactly for spheres centered on the origin.  The result depends only on
the quantity
$$R \equiv  H r \tau, \eqn\static $$
which is simply related to the static $\R$ coordinate by $R = \R - M$
\transform.
The expansions are
$$\theta_{out}= 2H + {2 R\over (R +M)^2},\qquad
\theta_{in}= 2H - {2 R\over (R +M)^2}. \eqn\inout    $$
It will be useful to define
$$ \alpha \equiv \sqrt{1+4MH} \eqn\alfa$$
and
$$  \beta \equiv \sqrt{1-4MH}. \eqn\betah$$
$\theta_{out}$ vanishes at
$$   R_{bh} = -{1\over 2H} \( 1+ 2MH - \alpha\) \eqn\rootbh $$
and
$$R_{dS} =-{1\over 2H}\left (1+2MH +\alpha \right), \eqn\rootdS $$
which correspond to the black hole and de Sitter horizons.
The ingoing expansion $\theta_{in}$
vanishes at
$$ R_{in} = {1\over 2H}\left ( 1-2MH -\beta \right ),
\eqn\apparent $$
which corresponds to the inner horizon.
The quantities $\alpha$ and $\beta$ will play an important role in what
follows.

In Fig. (5) we have drawn the coordinate patch
covered by $(r,\tau)$.
The coordinate patch is divided into
four regions, labeled I - IV, by various horizons. These horizons coincide
with the boundaries of the regions of trapped surfaces,
$R_{bh},\ R_{dS},\ R_{in}$, given in \rootbh\ -- \apparent.
Spheres in region I are
outer trapped, and the boundary between regions I and II corresponds to
$R_{dS}$.  Spheres in region III  are also outer trapped and
the boundary between regions II and III is given by $R_{bh}$.
Spheres in region IV are both inner  and outer trapped.

We now consider the two-black-hole solution where each mass is less than
the extremal limit $M_i<M_{ext}$.
The main
difference from the one-black-hole case, or the multiple black holes with
zero cosmological constant [\mp], is that these solutions are dynamical.
The apparent horizon will evolve in time.
Of course, with two black holes
an apparent horizon will no longer be precisely spherical, but it will be
approximately spherical
in limiting cases.
Hence we consider
$$\theta _{out} = 0 ,\eqn\exp$$
evaluated on appropriately centered spheres. (Trapped surfaces have been
found for initial data describing two uncharged black holes with $\Lambda > 0$
in [\nym].)

Spheres of sufficiently small radius, centered at $\vr_i$, are always trapped.
At early time, for $\tau \ll 0$, there is a solution for $\theta _{out} =0$
at $|\vec{r}-\vec{r}_i| =R_{bh} [M_i]/H\tau$,
since the equation is identical to
that of RNdS with mass $M_i$.
(Here $R_{bh}[M_i]$ is defined as in \rootbh, with $M$ replaced
by $M_i$.)
So, around each throat there is a region of outer trapped surfaces, surrounded
by an external region.
Actually, around each mass there is also
the second solution  to $\theta _{out} =0$ at $|\vec{r}-\vec{r}_i| =R_{dS}
[M_i]/H\tau$. These two solutions
correspond to the black hole horizon and the past de Sitter horizon in
the RNdS spacetime.

Spheres of constant $r$ (where $r$
is much larger than the coordinate distance between the two masses) resemble
spheres in the RNdS solution with mass equal to the total mass $M=\sum M_i$.
If $M<M_{ext}$, then at late times, as $\tau$ approaches zero, one again
has two (outer) apparent horizons at $r = R_{bh}/H\tau$ and $r=R_{dS}/H\tau$.
This suggests that the black holes coalesce. If $M>M_{ext}$, then there are
no apparent horizons at late time for large $r$.
We will have more to
say about this case in  Sec. 7.

One can understand the behavior of the apparent horizon as follows. For the
RNdS solution, the horizon is at fixed $R$, which corresponds to $r \propto
1/H\tau$. So when $\tau$ is large and negative, the horizon is at small $r$,
and when $\tau \approx 0 $, it is at large $r$. In both of these limits,
the two black hole solution resembles the one black hole solution (although
with different masses).

\chapter{Extensions}

\section{Locating the Horizons}

As one might expect from the RNdS solution,
the region of spacetime described by the metric \metric\ (with $   \tau, \vr$
taking all real values) is incomplete, even away from the curvature
singularity.
It is bounded by the analog of the de Sitter horizon at large $r$ and
the  inner black hole horizon and past white hole horizon at small
$|\vr -\vri|$. In both these regions the metric becomes approximately
spherically symmetric. We will establish the incompleteness by considering
radial null geodesics in these asymptotic regions.

We first derive
an equation for the affine parameter $s$ along a radial null geodesic
from the variational principle
$$ 
\delta \int \left(
-{1\over U^2}\left({d\tau\over ds}\right)^2 +
U^2 \left( dr\over ds\right)^2\right) ds = 0. \eqn\vary $$
Vary $\tau$ and use the outgoing null condition,
$$d\tau/dr = U^2, \eqn\nul $$
to find
$$ {d^2r\over ds^2} + 2H U\left({dr\over ds}\right)^2 = 0. \eqn\nulltwo$$
If a null geodesic is known as $\tau = \tau(r)$, the affine parameter
is given by quadratures,
$$ s = \int e^{2H \int U\left(\tau(r),r\right)dr} dr.\eqn\para$$

In the limit of large $r$, the function $U$ takes the simple form
$$ U \a H \tau + {M\over r}.   \eqn \ulr $$
where $M= \sum M_i$. In this case, the null condition \nul\ is
integrable when rewritten in terms of the variable $R$ of \static:
$$ {dR \over dr} = R + H(R+M)^2.  \eqn\outnull $$
We need only the asymptotic form of the solution.
If $M<M_{ext}$, then the right hand side has two roots at $R_{bh}$ and
$R_{dS}$.
Starting with $R>R_{bh}$ one finds that the solution always approaches $R_{dS}$
as $r \rightarrow \infty$ (see Fig. 6). Thus
$$U\left(\tau(r),r\right) \a {R_{dS} + M\over r}. \eqn\ulim$$
Using this in \para, and noting from \rootdS\ that
$R_{dS} + M = -(1+\alpha)/2H$ we find that
$$s\sim s_H + c r^{-\alpha} \eqn\affin $$
where $s_H$ is the horizon value of $s$, and $ c$ is a constant
of integration.
Thus $s$ remains finite as $r \a \infty$.
Asymptotically for large $r$, it follows that
$$ r \sim (s-s_H)^{-1/\alpha},  \quad
\tau \sim (s-s_H)^{1/\alpha}.\eqn\rxt$$
This not only shows that these quantities reach their horizon values at
a finite $s$, but also that $1/r$ and $\tau$ are not smooth functions
of $s$ (unless $1/\alpha$ happens to be integral).

For each mass $M_i$, we can find similar results for ingoing null geodesics
near the inner
horizon, in coordinates centered about $M_i$.
In the limit of small $r$, ingoing null geodesics satisfy
$${d\tau \over dr} = -U^2 =
- \left(H\tau + {M_i \over r} +
\sum_{j\ne i} {M_j\over r_j}\right)^2. \eqn\dtdr $$
The last term on the right is a constant which can be removed by shifting
the origin of $\tau$.
We then obtain an equation like \outnull\ whose solution has the limiting form
$$  R \a R_{in}[M_i], \qquad U\left(\tau(r),r\right)
\a {R_{in}[M_i] + M\over r}. \eqn\inlim$$
Eq. \para\ now implies that as $r$ goes to zero,
$$s\sim s_I + C r^{\beta_i} \eqn\zzz$$
where $s_I$ denotes the value of the affine parameter at the inner horizon,
$C$ is a constant of integration and $\beta_i \equiv \sqrt{1 + 4M_i |H|}.$
So the affine
parameter is again finite at the inner horizon. It follows that
$$ r \sim (s-s_I)^{1/\beta_i},  \quad
\tau \sim (s-s_I)^{-1/\beta_i}. \eqn\Rxt $$

\section{Extending Beyond the Horizons}

We now consider extensions across the horizons.
We begin by introducing new variables. For the single black hole, the
static $\R$ coordinate is good on the horizons. The closest analog for
the solutions \metric\ is $ R= Hr \tau $ introduced above, where $r=0$ is
chosen to correspond to the location of one of the masses.
We also set
$$y = \ln r, \qquad W = rU. \eqn\yandw $$
In $(R,  y)$ coordinates, the metric \metric\ takes the form
$$ ds^2 = -{(dR - Rdy)^2 \over H^2 W^2} + W^2 (dy^2 + d\Omega^2).
\eqn\Metric $$

We have seen that the metric approaches the solution
for a single black hole both
in the limit of large and small $r$.
However, in determining the behavior of the geometry across the horizon,
the {\it rate}
at which the metric approaches the single-black-hole solution is crucial.
We first show that all curvature scalars
remain finite as one approaches these horizons.
The $(R,y)$ part of the metric \Metric\ has
constant determinant. Thus the inverse metric has a similar form, with only
$W^2$ appearing in the denominator. A general curvature scalar will involve
terms consisting of derivatives of the metric and its inverse, multiplied by
powers of the metric and its inverse. All of these terms reduce to
derivatives of $W$ and $R$ divided by powers of
$W$. But all derivatives of $W$ remain bounded as
$y \a \pm \infty$, and since $W$ is finite on the horizon, these terms
cannot blow up. (By simply shifting the origin of $r$, one can apply this
argument to the horizons near each of the masses.)

It is tempting to conclude from this that the metric is smooth across the
horizon and can be analytically continued as in the single-black-hole case.
However this is incorrect. We will see that, in general, the horizon has
only finite differentiability. The curvature can even be singular at the
inner black hole horizon,
but the singularity is ``null,"
and so all curvature scalars remain finite\foot{The above argument
shows that all components of the Riemann tensor in $(R,y)$ coordinates stay
bounded, but this coordinate basis is not well behaved at the horizon.}.
To establish this result,
we will introduce coordinates that are good in a neighborhood of the
horizon. It will turn out that $R$ is a good coordinate but that
$r = e^y$ is not.
(We have already seen
in  \rxt\ and \Rxt\ that $r$ is not a smooth function of the affine
parameter along null geodesics.)
For  the single-mass solution, $W = R+M$, so the
metric \Metric\  depends only on $R$, and therefore is smooth.  However,
the effect of the
other masses is to modify the single-black-hole
metric by a power series in $r$, so the exact
metric is no longer smooth across the horizon.

The metric \Metric\ can be rewritten in the form
$$ ds^2 = {1\over H^2W^2} \[ dR + (HW^2 -R) dy\] \[ -dR + (HW^2 +R)dy\]
     + W^2 d\Omega^2. \eqn\factor $$
Our procedure for studying the extensions of the spacetime is to introduce
new coordinates $(u,v)$ in the neighborhood of each horizon, such that
the first quantity in brackets is proportional to $du$ and the second
is proportional to $dv$ at the horizon.

We first consider the de Sitter  horizon ($y\rightarrow\infty$).
If we choose the
origin of coordinates to be the center of mass,  the metric takes the form
\metric\  with
$$ U = H\tau + {M\over r } + {f(\theta,\phi)\over r^3 } + O(r^{-4})
\eqn\appro$$
where $M$ is the sum of the individual masses. In terms of $(R, y)$ coordinates
the metric takes the form \factor\ with
 $$W = R + M +
f(\theta,\phi) e^{-2y}+\cdots \eqn\w $$
We saw in Sec. 4.1 that $R$ approaches the
value $R_{dS}$ at the horizon. Expanding $HW^2$
about $R_{dS}$ yields
$$H W^2 \approx -R_{dS} - (1+\alpha)[R-R_{dS} + f e^{-2y}].\eqn\wsquare $$
Define new coordinates
$$ u = \[1+ {\aa\over 2R_{dS}}( R-R_{dS}) + {\aa(\aa +1) f\over 2R_{dS}
(\aa+2)}
e^{-2y} \] e^{-\aa y} \eqn\newcor$$
$$ v = -\[ R-R_{dS} + {(\aa +1) f\over \aa - 2} e^{-2y} \] e^{\aa y}. $$
The metric near the horizon becomes
$$ds^2 =  {2R_{dS} \over \aa H^2W^2}du dv+W^2 d\Omega^2. \eqn\Lead$$
So $u,v$  are good null coordinates near the horizon, which now corresponds
to $u=0$. The curves $v=$ constant cross the horizon. The metric \Lead\
depends on $W^2$ which
involves factors of $(R-R_{dS})$ and $e^{-2y}$. In addition, there are
corrections to
the leading order behavior \Lead\ that involve these same factors.
To express these terms
as functions of $u$ and $v$ we need to invert \newcor. Near the horizon
we have $u=e^{-\alpha y}$, and $R-R_{dS} = -[v e^{-\alpha y}+(\aa+1)
fe^{-2y}/(\alpha -2)]$.
But since $\aa = \sqrt{1-4M|H|} <1$,
as long as $v \ne 0 $ the second term is negligible compared to the
first as $y \a \infty$,
and we have $ R-R_{dS} = uv$. Thus corrections involving only powers of
$R-R_{dS}$ will be smooth at the horizon. However, $ e^{-y} = u^{1/\alpha}$
is not smooth in general.
Since the corrections start at order $e^{-2y} = u^{2/\aa}$ and $\aa <1$,
we see that the metric is always at least $C^2$ in these coordinates.
Since the metric is $C^2$, there is no curvature singularity. But one can
show that in general, certain components of derivatives of the curvature
diverge at the horizon, so there are no coordinates for which the metric
is analytic. As a result, the extension across the horizon is not unique.
One can match onto essentially any solution of the form \metric\ with
the same total mass.
(Similar behavior of finite differentiability across a horizon in an
exact solution was found in [\piotr].)

The differentiability at the horizon
can be increased in two ways. If the total mass
$M$ is chosen so that $1/\aa$ is an integer $n$, then $e^{-y}$ is smooth and
the
metric is $C^\infty$.
This occurs when
$$4 M |H|= 1-{1 \over n^2 }. \eqn\disc $$
For these values, the smooth continuation consists of matching the
spacetime onto one with the same positions and magnitudes of all the masses
so that all multipole moments agree, but with the opposite  sign of $H$.
We do not understand the physical significance of these special masses.

The second way to increase differentiability is to arrange the masses so
that their first $n$ multipole moments vanish. Then the perturbation will
begin at $e^{-ny}$. These solutions may provide a simple model of smoothness
of null infinity in asymptotically flat spacetimes.
It has been suggested [\schmidt]
that the behavior of fields at null infinity may depend on their fall-off
near spatial infinity. If the unconstrained part of the
initial data falls off more quickly in spacelike
directions, then
perhaps the evolved fields will be more differentiable
at null infinity. This is very similar to
the behavior we find in the multi-black-hole
solutions.

We now consider the inner horizon associated with one
mass $M_i$ which we choose to be
located at the origin.
Near $r=0$ we have
$$U = H\tau + {M_i \over r} + c + g(\theta,\phi) r + \cdots \eqn\approxtwo$$
where $c$ is a constant that can be removed by changing the origin of $\tau$.
Thus
$ W = R + M_i + gr^2 +\cdots$.
We showed in Sec. 4.1 that
$R$ approaches the constant $R_{in}[M_i]$ at the inner horizon.
Expanding $HW^2$ near $R=R_{in}[M_i]$ yields:
$$ HW^2 \approx R_{in}[M_i]+ (1-\b_i) \(R-R_{in}[M_i] + g
e^{2y}\) \eqn\W $$
where, as before,  $\b_i = \sqrt{1+4M_i |H|}$.
Define new coordinates by
$$  u = \[R-R_{in}[M_i] + \({ 1- \b_i \over 2-\b_i}\)g e^{2y}\] e^{-\b_i y} $$
$$  v =\[1-{\b_i \over 2 R_{in}[M_i]}\(R-R_{in}[M_i]\)+{\b_i(1- \b_i)  \over
2 R_{in}[M_i]
(\b_i+2)}
g e^{2y}\] e^{\b_i y} \eqn\newcor $$
Then the leading order behavior of the metric near the horizon takes the
simple form
$$ ds^2 = \({2 R_{in}[M_i] \over \b_i H^2}\) {du dv\over W^2}
+W^2 d\Omega^2 \eqn\lead $$
(The leading factor in parentheses is just a constant.)
So $u,v$  are again
good null coordinates near the horizon, which now corresponds
to $v=0$.
Near the horizon
we have $v=e^{\b_i y}$, and $R-R_{in} = u e^{\b_i y}+ (\b_i
-1)ge^{2y}/(2-\b_i)$.
But since
$\b_i = \sqrt{1+4M_i|H|}$, every black hole with $M_i<M_{ext}$
has $\b_i<2$. Thus
as long as $u \ne 0 $ the second term is negligible compared to the
first near the horizon,
and we have $ R-R_{in}[M_i] = uv$. Corrections involving only powers of
$R-R_{in}[M_i]$ will again be smooth at the horizon.
However, $r = e^{y} = v^{1/\b_i}$,
and $1 < \b_i < 2$. Thus
$r$ is not smooth at the inner horizon.
Since the corrections start at order $r^2$,
the metric is $C^1$ but not $C^2$ in these coordinates.

To see that there are no better coordinates for which the metric is smooth,
we compute a component of the curvature. Let $l = \d/\d v$
and $\eta = \d/\d \theta$. Then
$R_{v\theta v \theta} = R_{\mu\nu\rho\sigma} l^\mu \eta^\nu l^\rho \eta^\sigma$
contains several terms
that are finite at the horizon. But it contains one term which is infinite
there. This is $g_{\theta \theta, vv}$, which involves two derivatives of
$v^{2/\b_i}$. This divergence is null since it only occurs at $v=0$.
We will return to the physical interpretation of  this singularity shortly.

Unlike the case of the de Sitter horizon, there are no special values
of the mass for which the inner horizon becomes smooth. However one
can still increase the differentiability by
carefully arranging the other masses so that
the first $n$ powers of $r$ cancel in the expansion of $U$ in \approxtwo.
In particular, one could remove the
curvature singularity this way.  However, the inner horizon associated with
the other black holes will still be only $C^1$.

It is interesting to compare this situation with the case of
zero cosmological constant. There, it was shown [\hh] that the spacetime
describing several $Q=M$ black holes has an analytic extension across each
inner horizon, which simply corresponds to letting $r$ become negative.
In other words, when $\Lambda = 0$,
$r$ is smooth at the horizon and there
is no analog of the singularity we find above. This is consistent with our
results since $\Lambda = 0$ implies $\beta_i = 1$ for all $i$,
so $r=v$ at the inner horizon.

The remaining horizon is the past white hole horizon $r=0, \ \tau = -\infty$.
One can show that the situation here is similar to the de Sitter horizon.
The metric is always $C^2$ and can be made $C^\infty$ if the individual
mass $M_i$ takes one of the discrete values \disc\ . Notice that if more than
one mass has one of the discrete values, the total mass exceeds the extremal
limit \extremal.

\section{Physical Interpretation of the Lack of Smoothness}

We now consider the physical interpretation of the lack of smoothness we
have found at the horizons. Consider the solution with two black holes
and take the limit where one mass becomes much less than the other.
In this limit, we can think of the small mass as a test particle moving in
the background of the large black hole. The single black hole solution is,
of course, smooth everywhere. But the perturbation in the metric and Maxwell
field obtained in this limit involves a function of $r$ (or $1/r$), which is
not
smooth at the horizon. This seems rather unphysical.
We would expect a $q=m$ test particle
to radiate as it falls into the black hole, and the field it produces
should remain smooth at the de Sitter horizon. This can be made more
precise in terms of initial data. To avoid singularities associated with
point particles, we can model the test particle by a small ball of $q=m$
dust. Let us take our initial surface to be the $T=0$ surface in the RNdS
metric.
One can certainly find smooth initial data for the
linearized Einstein-Maxwell field equations that satisfy the constraints
with the ball of dust as a source. The evolution of these initial data
must be smooth everywhere
in the domain of dependence, which includes both the de Sitter horizon
and past white hole horizon. The fact that the multi-black-hole solutions
are not smooth at these horizons can be interpreted as saying that they
describe more than colliding black holes. In addition, they contain a
distribution of electromagnetic and gravitational radiation which is not smooth
everywhere in the spacetime.
We expect that there are other (undoubtedly more complicated) solutions
describing colliding black holes that do not suffer from this lack of
smoothness.

The singularity at the inner horizon is qualitatively different. This is
because it
is on the boundary of the domain of dependence of the initial data surface
described above. The behavior we find
is reminiscent of the instability of the inner black hole horizon
which has been extensively studied for $\Lambda = 0$ [\cauchy]. (When
$\Lambda >0$ it has been argued that the inner horizon might be
stable [\dscauchy].) A key difference, however, is that the
previous analyses were based on a perturbation
expansion, while the solution \metric\ treats the effect of the other masses
exactly. One unusual feature of these solutions is that only half of the
Cauchy horizon becomes singular. The other half lies inside the $(\vr, \tau)$
coordinates and is $C^\infty$. Perhaps this is related to the extra
radiation in the spacetime that is responsible for the lack of smoothness
at the de Sitter horizon.

The idea that the finite differentiability at the de Sitter and past white hole
horizons is a result of additional radiation is supported by the fact
that one can construct exact initial data for multiple charged black holes with
$Q<M$. These initial data are smooth at the white hole horizon.
This construction relies on the $\Lambda = 0$
initial value solutions
of Brill and Lindquist [\bl] for time-symmetric, arbitrarily placed wormholes
of general mass and charge, and on the method of Nakao et al [\nym]
to turn such a solution into one with a cosmological constant. The 3-metric
and electromagnetic potential of [\bl] are
$$ds^2 = (\chi \psi)^2 (dx^2 + dy^2 + dz^2) \qquad A = \nu \ln(\chi/\psi)
\eqn\iv $$
$$\chi = 1 + \sum_i a_i/r_i, \qquad \psi = c + \sum_i b_i/r_i,$$
where $\nu$ is a unit normal form to the initial surface, and $a$,
$b$ and $c$ determine the wormholes' masses and charges (approximately,
$M_i \approx a_i/c + b_i$, $Q_i \approx -a_i/c + b_i$).
Eq \iv\  satisfies the $\Lambda = 0$ Einstein constraints if the extrinsic
curvature $K_{ij}$ vanishes; therefore it satisfies the $\Lambda \neq 0$
constraints if we put [\nym]
$$K_{ij} = \pm H g_{ij}, \qquad H = \sqrt{\Lambda/3}. \eqn\extrinsic$$

Special cases of \iv\ are initial values of the known dynamical solutions, as
follows:
The general $Q\ne M$ RNdS geometry in cosmological
coordinates [\bh],
$$ ds^2 = -{1-{M^2 - Q^2 \over 4 e^{2Ht} r^2}\over U^2}dt^2 +
e^{2Ht}U^2(dr^2 + r^2 d\Omega^2) $$
$$= -{\left(1 - {M^2-Q^2\over 4R^2}\right)
(dR - Rdy)^2 \over H^2W^2} + W^2(dy^2 +d\Omega^2) \eqn\iso $$
$$U = 1 + {M \over e^{Ht} r} + {M^2 - Q^2 \over 4 e^{2Ht} r^2}
\qquad W = R + M + {M^2-Q^2 \over 4R^2} $$
has the form \iv\  with $a_1 = (M-Q)/2$, $b_1 = (M+Q)/2$,
$a_i = 0 = b_i$ for $i \neq 1$, $c = 1$, on the surface $t=0$.
The solutions \metric\ have the form \iv\  with
$a_i = 0$, $b_i = m_i$, and $c=H\tau $.

We also note that the geometry of the surface $\tau = 0$ of the solution
\metric,
as well as that of the surface $\R = M \pm \sqrt{M^2-Q^2}$ in the general
RNdS geometry (in {\it static} coordinates) can be generalized by
initial values of the type \iv\  with $c = 0$.

\chapter{Event Horizons}

All of the extensions considered above have a $\Im^+$
beyond the ``de Sitter horizon," which corresponds to the
limit $(r, \tau) \rightarrow (\infty, 0)$, but $R \rightarrow
R_{dS}$ finite, in our original coordinates \metric.
In these original coordinates we can therefore identify
null curves that can go to $\Im^+$ as those that reach
this horizon. By contrast, the black hole horizon as defined in
the Introduction (see also
Ref. \snkm) is contained entirely within the
original chart: it is the boundary of the region of events
that can be causally connected to the de Sitter horizon.
Thus we identify events from which outgoing null geodesics
must reach the $U=0$ singularity as lying inside the black
hole horizon.

In this section we establish several properties of null geodesics that
have implications for the black hole horizon. For simplicity we
confine attention to the  solution \metric\
for two centers with identical
mass parameter $M/2$ with $M<M_{ext}$,
separated in the 3D Euclidean base space by
distance $2a$. The origin of the Euclidean coordinate system  is at the
midpoint between the two centers. The Euclidean line connecting
the two will be called the axis, and the perpendicular plane at
the origin is the midplane. We show that for each sufficiently
late (but negative) time $\tau$ there is a sphere of radius $r(\tau)$
such that all outgoing null geodesics from the sphere will reach the
$U = 0$ singularity --- hence no causal curve from inside can
reach $r=\infty$; but that
there are points outside the sphere
from which causal curves reach $r=\infty$ in finite affine parameter.
That is, the sphere lies within the event horizon. We also
show that at sufficiently early times, all points in the midplane
can be connected to $r = \infty$ by causal curves; in other words,
at early times the horizon does not meet the midplane.
Finally, we will show that at sufficiently early times  the event horizon
consists of a sphere centered on each mass.
Taken together these results prove that
the black holes coalesce. The
event horizon at early times has two disconnected pieces, but at late
times it has only one.

We first consider the region inside the horizon. Every outgoing null
geodesic must satisfy
$$U^2 dr^2 \le  {d\tau^2 \over U^2}.
\eqn\late $$
A lower limit on the ``potential" $U$ is obtained by pretending
that the total mass $M$ is concentrated at the greatest possible
distance, $r+a$, so that
$${d\tau \over dr} \ge U^2 > \left(H\tau + {M \over r+a}\right)^2.
\eqn\more $$
In terms of new variables $R_\star = H \tau(r+a) $, $y_\star = \ln(r+a)$
the inequality simplifies,
$${dR_\star \over dy_\star} < R_\star  + H(R_\star + M)^2
\equiv H( R_\star - R_{dS})(R_\star - R_{bh}). \eqn\less $$
The slope of the function on the RHS is positive at its smaller zero,
$R_{bh}$ (Fig. 6).
Suppose  that initially $R_\star < R_{bh}$,
i.e., $a< r < (R_{bh}/H\tau) - a$.
This can be fulfilled if $\tau > -R_{bh}/2a|H|$.
Then $R_\star$ decreases with $r$ and becomes
negative at a finite $r$. But then $\tau$ is  positive, and
as it increases further with $r$,
the singularity at $U =  0$ is reached at a finite $r$.
Thus, as $\tau$ approaches zero (from below), there exist spheres of constant
$r$  enclosing both masses
such that all outgoing null geodesics hit the singularity.

We now show that at early times, a similar result holds for small spheres
centered on each mass. For simplicity,
let us shift the origin  of
spherical coordinates to be at one of the masses, and assume
$r < a$. Then the ``potential" from the other mass is at least
$M/6a$, so that instead of Eq~\more\  we have
$${d\tau \over dr} \ge U^2 > \left[H\left(\tau+{M \over 6aH}\right)
 + {M \over 2r}\right]^2.  \eqn\bounddtdr$$
Introducing the new variable $R_+ = Hr(\tau+{M \over 6aH})$ leads
to an equation analogous to \less,
$$dR_+/dy < R_+  + H(R_+
+{\scriptstyle {1 \over 2}} M)^2. \eqn\str $$
Analyzing  \str\ in the same way as  \less\ we then find that
events satisfying $R_+<R_{bh}[M/2]$ are inside the horizon.
This inequality is satisfied at any early (negative) $\tau$
by all sufficiently small $r < R_{bh}[M/2]/(H\tau + M/6a)$.

To explore the outside of the black hole horizon, we ask
under which conditions some causal curves from a given
event can reach $r = \infty$.
We first confine attention to the midplane, and obtain an
upper limit on the inverse speed of purely radial null curves:
$${d\tau \over dr} = U^2 = \left(H\tau +
{M \over \sqrt{r^2+a^2} }\right)^{2} <
\left(H\tau + {\sqrt{2}M \over r+a}\right)^2.\eqn\dtdrtwo$$
In terms of the variable $R_\star,~y_\star$ introduced in Eq \less\
this becomes
$$dR_\star / dy_\star > R_\star  + H(R_\star + \sqrt{2} M)^2. \eqn\pls $$
Arguing as before (see Fig. 6) we find that if
$R_\star > R_{bh}[\sqrt{2}M]$ is satisfied  initially
for some $r$, then $R_\star$ will
stay positive for all larger $r$. But this
means that $\tau$ will remain negative as $r$ increases, avoiding the
singularity.

For example, if the null curve starts at any $r$, but
with  $\tau <  R_{bh}[\sqrt{2}M]/Ha$, it satisfies the
initial inequality and hence
will be able to escape to $r=\infty$.  Thus at early times no
point in the midplane is within the horizon (provided of course
that a real $R_{bh}[\sqrt{2}M]$ exists; if $M$ is close to the extremal limit,
a more accurate investigation of the null geodesics is needed).

Similarly, for any $\tau < 0$
we can find a sufficiently large
$$r> (R_{bh}[\sqrt{2}M]/H\tau)-a\eqn\suffic$$
 so that $R_\star > R_{bh}[\sqrt{2}M]$ and null geodesics can escape to
 infinity.
In fact, for sufficiently large
$r$ the argument leading to the increase of $R_\star$ is valid also for points
not on the midplane. Thus outside the sphere of events that must causally
lead to the singularity there are events from which causal escape
to infinity is possible. At early times, therefore, the event horizon
surrounds each mass separately, and expands approximately
according to $r = R_{bh}[M/2]/H\tau$. At $\tau \approx R_{bh}/2aH$
the event horizon enters the midplane, and thereby changes its topology
to a surface surrounding both centers.

\chapter{Charged Shells and Dust}

As stated in the Introduction, in our test of cosmic censorship,
 we will need to introduce
a shell of charged dust to remove an unwanted singularity. The dynamics
of these shells is also of interest in its own right. For example,
one can ask whether it is even possible to form $Q=M$ RNdS black holes from
collapsing matter.  For $\Lambda=0$, Boulware [\boulware] studied the
dynamics of charged shells.  He found that a shell
having charge density equal to its rest mass density and also equal to its
total
mass density does not collapse.
Rather, as one might expect,
it stays at constant area.  For $\Lambda >0$, on the other
hand, we find that a simple extension of Boulware's calculation does give
collapse to form $Q=M$ RNdS black holes.

\section{Shells}
Consider a spherical charged shell in an otherwise empty spacetime with
$\Lambda>0$.  Inside the shell Birkhoff's theorem guarantees that
the metric is de Sitter.  Outside the shell, we match the
metric to RNdS.  In order to determine the motion of the shell, one integrates
Einstein's equation across the shell to obtain jump conditions on the
curvature. Working in static coordinates, one finds that
a shell having charge
$Q$ equal to its rest mass $M$ also equal to its total mass ({\it i.e.} no
kinetic energy) follows the same path \testpath\ as the radially moving $q=m$
test particle discussed in Sec. 2.3.
In cosmological coordinates then, the shells stay at constant comoving
radius.

The physical picture of shell collapse is then quite simple.
The metric has the form \metric, with $U=H\tau$ for $r<r_s$ and
$U=H\tau+M/r$ for $r>r_s$.
The comoving radius of the horizon
however, changes with time.  For $M<M_{ext}$, there is a horizon  at
$R = Hr_H\tau = R_{bh}$.
For large negative $\tau$, $r_H$ is inside the shell and hence does not
correspond to a horizon.  As $\tau$ increases
towards zero,  $r_H$ sweeps past the shell, which is now contained within
the black hole.  The same process can be used for a number
of $Q=M$ shells to form the multi-black-hole solutions, if the
shells are located at an equipotential of $U$ and have a mass
and charge distribution appropriate to a constant $U$ in the interior.
Then the metric is de Sitter inside each shell,
and the multi-black-hole metric outside.
 All the shells stay at constant comoving radius, and eventually become
black holes which later merge.
Spacetime diagrams for a
single collapsing shell are given in Figs. (7a,b) for undermassive and
overmassive shells.  The shells take a finite cosmological time to reach the
singularities, but an infinite proper time.

The shells in Fig. (7) all start on the right side of the Penrose
diagram and collapse to the singularity on the left.
We can also introduce shells on the left which
collapse to the singularity on the right. This follows from the $T \a -T$
symmetry of the RNdS solution in static coordinates.
The motion of these shells is simple in the
$(r^\prime,\tau ^\prime)$ coordinates defined in section 2.3.
The shells stay at constant $r^\prime$.
For $r^\prime<r^\prime_s$ the metric will be de Sitter, and for
$r^\prime >r^\prime_s$ the metric will be RNdS.
These shells will be useful for us in Sec. 7 in constructing our potential
counterexample to cosmic censorship.

\section{dust}

It turns out to be easy to handle arbitrary configurations
of $q=m$ dust
as well.  Consider a general metric  and gauge field of the form
$$ ds^2= {1\over U^2}d\tau^2 +U^2 d\vr\cdot d\vr,\qquad
A_\tau ={1\over U},\eqn\general  $$
where $\partial U/\partial \tau=H$.  For matter consider charged,
comoving dust.
The matter stress-energy and current density are given by
$$T_{ab}^{dust}=\rd u_au_b, \qquad J_a=\rc u_a,\eqn\stress $$
where $u^a= {1\over U}({\partial\over\partial t})^a$ is the four velocity of
the dust.
The Hamiltonian constraint is given by
$$ \eqalign{{\cal H} & = - ^{(3)} R +{1\over g}
\left(\pi^{ij}\pi_{ij}-\half\pi^2\right ) \cr & =
4{\nabla^2 U\over U^3} - {2\over U^4}\delta^{ij}\partial_iU\partial_jU -6H^2\cr
& = - 16\pi(\rho_{dust}+\rho_{Maxwell}+\rho_{cosmo}).}\eqn\constraint$$
The third term on the second line is just the cosmological term.
The second is equal to
the energy density of the Maxwell field. Thus,
the first term must be compensated
by the energy density of the dust.  For a solution then, one requires
$$\nabla^2 U = -4\pi \rd U^3. \eqn\dust $$
The Maxwell constraint gives the same equation \dust , if $\rc =\rd$.
The evolution equations require that $T^{dust}_{ij}=0$ and that
$\partial_\tau (U^3\rd) = 0$.  The first requirement is satisfied for comoving
dust.  The second condition requires that the function $\rt(\vx)=U^3\rd$
be independent of $\tau$.  The constraint equation \dust\ is then
$$\nabla^2 U =-4\pi\rt. \eqn\poisson$$

We see that we can freely specify the function $\rt$.
This, together with $\partial U/\partial \tau=H$, and the boundary
condition $U \a H\tau$ as $r \a \infty$ determine a unique $U$.
The physical dust
density $\rd$ is then derived from $\rt$ and $U$.
Note that the volume element $\sqrt{h} = U^3$ for the spatial
metric, so that the total mass of a dust cloud is constant and given by
$$M=\int d^3x \sqrt{h} \rd = \int d^3 x \rt.\eqn\mass $$
Other measures of the dust cloud, e.g. its volume, are of course
time dependent and consistent with the picture of a collapsing cloud.

\chapter{Testing Cosmic Censorship}

Having developed the necessary properties of the  multi-black-hole solution
we now turn to our test of cosmic censorship. Consider
two black holes, each with mass  less than
the extremal limit, but whose sum is greater.
Since the multi-black-hole solutions are not spherically symmetric,
it is difficult to describe their global structure
completely.
It is therefore convenient to focus on
a two dimensional slice of the spacetime.  Consider first a curve
which comes up one throat and down the second on a constant $\tau$ surface.
Evolving in time, we obtain a two dimensional subset of the spacetime
which is nonsingular for $\tau < 0$, but develops a timelike singularity
between the two throats at a certain time $\tau_0 >0$ (when $U=0$).
This appears to be a violation of cosmic censorship in which a naked
singularity
evolves from regular initial conditions. However, since this subset of the
spacetime does not include the asymptotic region, one does not know if the
singularity is hidden behind an event horizon.

A better choice is to consider the curve which starts at infinity and goes
down one of the throats on a constant $\tau$ surface. Evolving in time,
we obtain a two dimensional slice of the spacetime whose
Penrose diagram
is shown in Fig. (8a). For small $r$, the solution looks like a
single subcritical black hole, except that $\Im^+$ has been pushed off to the
future of the region shown. For large $r$,  it resembles the overmassive
case with its singularity.
This singularity is naked and exists for all time.  It
is independent of
whether the two black holes have collided or not, and is just a reflection
of the fact that the total mass  always was greater than the extremal limit.
Fortunately, this singularity can be removed by adding a shell of charged
matter.
This follows from the fact that the solution, at large $r$, reduces to the
overmassive RNdS solution, and we saw in Sec. 6.1 that the singularity in this
solution could be removed by a shell. The Penrose diagram for the solution
with the shell added in shown in Fig. (8b).

In the absence of a cosmological constant, one usually requires that
a counterexample to cosmic censorship have nonsingular data on a surface
that is asymptotically flat outside of a compact set. In the presence
of a cosmological constant, the analog would be nonsingular data on a
compact manifold. Our example does
not have a compact surface, but it has  what might be considered
``the next best thing."
Consider the surface $S$ shown in Fig. (8b). This
surface is defined by
$\tau$ equal to a negative constant outside the shell, and any spacelike
surface
inside the shell that continuously joins to it. The initial data on this
surface is nonsingular everywhere. The surface is not compact, but has
two infinite throats (only one of which is shown on the figure). However,
each of these throats is surrounded by a trapped surface. So one would
not expect that  the asymptotic regions down the throats
could influence the solution in
the interior.

The initial data on $S$ uniquely determines the solution up to the
Cauchy horizon. As we have discussed in Sec. 4,
the solution past the horizon is not
unique but all extensions have a curvature singularity shown at the right
in Fig. (8b).
It is clear from the diagram that all observers originating from $\Im^-$
reach the Cauchy horizon, and if they extend beyond, they will see the
singularity.
We have also seen in Sec. 4 that the horizon is not smooth.
Generically, for the
class of solutions \metric,
the horizon is $C^1$ but not $C^2$. If we end the spacetime
at the horizon then, of course, there is no violation of cosmic censorship.
But one can find choices of parameters such that
the  Cauchy horizon associated with one of
the masses is at least $C^2$ and there is
no curvature singularity. Furthermore, even in the general solution,
geodesics can be extended beyond the horizon.
So it seems reasonable to conclude that
cosmic censorship is violated in these examples.

It is clear that this violation is
associated with the infinite throats in the initial data. One way
to see this is that in cases when the horizon is $C^2$ or smoother, it is
homogeneous. There is no point where the curvature is becoming large,
associated
with the beginning of a naked singularity. The singularity seems to ``come
in from infinity."
The fact that the asymptotic regions are hidden behind trapped
surfaces in the initial data, does not seem to be sufficient to prevent
the violation of cosmic censorship. In a sense, all of space
collapses down the throats carrying all observers with it.

In light of this, it is natural to ask whether cosmic censorship would
be violated if one first formed the black holes from regular initial
conditions.
We have shown that
one can, in fact, form the black holes using shells of charged dust. If one
also removes the singularity at infinity with another shell as we have
discussed above, one has
compact initial data. The problem now is that the two types of shells
collide in the spacetime before any singularities have formed.
We do not know the solution explicitly after this occurs.
However, even if naked singularities
were found later in the evolution, one would not know whether they
were fundamental, or artifacts of the dust approximation. It is well
known that naked singularities can form in spherically symmetric dust
collapse. These shell crossing or shell focusing singularities can also
occur in the absence of gravity and hence have nothing to do with cosmic
censorship.

\chapter{ Conclusions}

Motivated by a new test of cosmic censorship, we have studied the global
structure of the multi-black-hole solutions \metric. We have found that,
if the total mass is less than the extremal limit \extremal, then
they describe black holes which coalesce. It is remarkable that an analytic
solution describing coalescing black holes can be expressed in such a simple
form. Somewhat surprisingly, we have
also found that these solutions contain radiation which is not smooth at
the de Sitter (and past white hole) horizon. Perhaps the presence of this
radiation is related to the simplicity of the solution.

The test of cosmic censorship was based on the fact that there is an
upper limit to the mass of a black hole  when the cosmological constant
is positive. We have seen that colliding two black holes which are each
less than the extremal mass, but whose sum is greater, does produce naked
singularities.

However, we cannot yet claim that this is a serious violation of cosmic
censorship for two reasons. The first concerns how generic the violation is.
As we have seen, the
most general of the exact solutions has the naked singularity protected by
a Cauchy horizon with a weak singularity. However, the exact solutions
only describe a subset of black hole collisions with a positive
cosmological constant. The initial position and masses can be specified
arbitrarily but
not their initial velocities. And, of course, one cannot specify arbitrary
gravitational and electromagnetic radiation. It is not clear whether the
most general solution has a Cauchy horizon with a stronger singularity.
If so, then cosmic censorship would be preserved.

The other reason concerns the fact that the example involves eternal black
holes and not ones which formed from compact initial conditions. It is not
yet clear how physically reasonable collapse would affect the formation of
naked singularities.

\vskip 1cm

\centerline{\bf Acknowledgments}
It is a pleasure to thank P. Chrusciel, J. Isenberg, V. Moncrief, and
B. Schmidt
for discussions.
This work was supported in part by NSF Grants PHY-8904035,
PHY-9008502 and NSF-THY-8714-684-A01.  JT and DK thank
the Institute for Theoretical Physics for its hospitality and support.

\vskip 1cm

\centerline{Figure Captions}

Fig.~1: Penrose conformal diagram for the $Q=M$ RNdS
geometry with $M<M_{ext}$. The maximally extended spacetime continues
indefinitely in all directions. The region covered by the cosmological
coordinates $(\vr, \tau)$ lies inside the bold lines.
Two horizons are labeled. The inner horizon is the extension
of the line labeled $r = \infty$ to the region between the singularities
(also see Fig.~5).
The solid curves represent $\tau =$ constant surfaces and the
dotted curve shows a typical $r =$ constant surface.
The dashed line denotes $\tau = 0$.

Fig.~2: Penrose diagram for the $Q=M$ RNdS geometry with $M=M_{ext}$.
The notation
is the same as for Fig.~1.

Fig.~3: Penrose diagram for the $Q=M$ RNdS geometry with $M>M_{ext}$. The
notation
is the same as for Fig.~1.

Fig.~4: Qualitative representation of the geometry on spacelike surfaces
of constant $\tau$: (a) $\tau < 0$, (b) $\tau = 0$,
(c) $0 < \tau \ll \sum M_i/Ha$, where $a$ is the typical coordinate
distance between the centers, (d) $M_i/Ha \ll \tau$.

Fig.~5: Regions of trapped surfaces and the horizons that separate them
for the RNdS geometry.

Fig.~6: A plot of the function $F(R) = R + H(R+M)^2$. In determining the
motion of radial null geodesics, one is lead to inequalities of the form
$dR/dy < F(R)$ or $dR/dy > F(R)$. In the former case, it is clear that
if $R<R_{bh}$ initially, it will continue to decrease  and become negative.
These curves must hit the singularity. In the latter case, if $R_{bh} < R <
R_{dS}$ initially, then it must remain positive. These curves reach the
de Sitter horizon. If equality holds, then $R$ approaches $R_{dS}$.

Fig.~7: The motion of a $Q=M$ shell in a RNdS geometry, (a) undermassive, (b)
overmassive case. The shell follows a curve
$r =$ constant $= r_0$. In the exterior region,
$r>r_0$ (unshaded part of the diagram) the
geometry is RNdS. In the interior (shaded, including the origin $r = 0$)
spacetime is homogeneous de Sitter space. The shell eliminates the singularity
that would otherwise be present in the right half of the digram (cf.~Figs
1 and 3). Shells that replace the singularity on the left by
a de Sitter interior follow a curve $r' =$ constant. The corresponding Penrose
diagrams would be  reflections of (a) and (b) about a vertical axis.

Fig.~8: (a) Penrose diagram for a two dimensional subspace of the
two black hole solution.
 Each hole is undermassive, but their sum is
overmassive. Near $r=\infty$, the left part of the diagram is similar
to Fig.~3  (single mass with $M>M_{ext}$), and near $r=0$ the right half
resembles Fig.~1 (single mass with $M<M_{ext}$). (b) Similar to (a) but with
a shell replacing the singularity on the left.

\refout
\end